\documentclass{ws-procs975x65}

\newcommand{\be}{\begin{equation}}
\newcommand{\ee}{\end{equation}}
\def\dmf{\dot{\mathfrak{M}}}
\def\sun{\hbox{$\odot$}}
\def\degr{\hbox{$^\circ$}}
\begin{document}

\title{Formation and appearance of pulsar-like white dwarfs}

\author{N. R. IKHSANOV$^*$ and N. G. BESKROVNAYA }

\address{Pulkovo Observatory, Saint-Petersburg, 196140, Russia\\
$^*$E-mail: ikhsanov@gao.spb.ru}

\begin{abstract}
Accretion-driven spin-up of a magnetized white dwarf in a close binary system is
discussed. We address a situation in which the magnetic field of the white dwarf is
screening during the accretion phase and re-generating due to the field diffusion
through the accreted material after it. We find this scenario to be effective for a
formation of massive pulsar-like white dwarfs.
\end{abstract}


\bodymatter

 \section{Three basic states of a compact star}

A magnetized, rotating compact star can be observed in one of the following three basic
states\cite{Shvartsman-1970}: {\it ejector} (spin-powered pulsar), {\it propeller}, and
{\it accretor}. All of these states are observationally confirmed for neutron stars (see
Table\,\ref{tb1}). The white dwarfs in the accretor state are observed in Cataclysmic
Variables (CVs). The low and moderately magnetized white dwarfs (with the surface field
$B \leq 1$\,MG) in interacting close binary systems usually accrete material from a
Keplerian disk while strongly magnetized ($B_* \sim 10 - 1000$\,MG) white dwarfs, which
are refereed to as Polars, accrete material from an accretion channel\cite{Warner-1995}.
The only white dwarf resembling a spin-powered pulsar is currently identified with the
degenerate companion of a peculiar CV AE~Aquarii\cite{Ikhsanov-etal-2004,
Terada-etal-2008, Ikhsanov-Beskrovnaya-2012}.

 \begin{table}
 \tbl{Three basic states of a magnetized, rotating compact star}
 {\begin{tabular}{lclc}
     \toprule
  State &
  Association &
  Energy release rate  &
  Observed luminosity \\
     \colrule
  Ejector &
  Spin-powered Pulsars  &
  $L_{\rm md} \simeq f_{\rm m} \dfrac{\mu^2 \omega_{\rm s}^4}{c^3}$ &
  $L_{\rm md} > L_{\rm obs}$ \\
     \noalign{\smallskip}
  Propeller &
  X-ray transients &
  $L_{\rm pr} \leq  \dmf\ v_{\rm esc}^2(r_{\rm m})$ &
  $L_{\rm md} < L_{\rm pr} \leq  L_{\rm a}(r_{\rm m})$, \\
      \noalign{\smallskip}
 Accretor &
 Accretion-powered Pulsars &
 $L_{\rm a} \simeq  \dmf_{\rm a} \dfrac{GM_*}{R_*}$ &
 $L_{\rm sd} < L_{\rm pr} < L_{\rm a}(R_*)$ \\
  \botrule
 \end{tabular}
 }
 \begin{tabnote}
$L_{\rm sd}$ and $L_{\rm pr}$ are the spin-down power of an Ejector and Propeller, respectively;
$L_{\rm a}$ is the accretion power; $\mu$ is the dipole magnetic moment, $\omega_{\rm s}$ is the
angular velocity, $M_*$ is the mass and $R_*$ is the radius of the compact star; $\dmf$ is the
mass capture rate and $r_{\rm m}$ is the radius of the magnetosphere of the compact star;
$v_{\rm esc}(r) = \sqrt{2GM_*/r}$ is the escape velocity at a radius $r$; $f_{\rm m}\sim 1-4$
is a dimensionless parameter\cite{Beskin-2010}.  \\
\end{tabnote}
\label{tb1}
\end{table}

 \section{White dwarfs in the ejector state}

Observations show strong magnetization\cite{Wickramasinghe-Ferrario-2000} (up to a
few\,$\times 10^9$\,G) and fast rotation\cite{Sion-1999} (up to a period of 30\,s) of
white dwarfs not to be very unusual. This indicates that an existence of a strongly
magnetized fast rotating white dwarf cannot be excluded. One of these objects (a white
dwarf with the spin period of 33\,s and the surface field of $\sim 50$\,MG) is
discovered in AE~Aquarii. The spin-down power of this white dwarf exceeds its bolometric
luminosity by a factor of 300 and can be explained in terms of the spin-powered pulsar
energy-loss mechanism provided the dipolar magnetic moment of the star
is\cite{Ikhsanov-1998} $\sim 10^{34}\,{\rm G\,cm^3}$. The fast rotating extended
magnetosphere in this case prevents the surrounding material from reaching the stellar
surface. The interaction between the surrounding material and the stellar magnetic field
leads to a formation of a stream which is flowing out from the binary
system\cite{Ikhsanov-etal-2004}. The spin-down power of the white dwarf is released
predominantly in a form of the wind of relativistic particles\cite{Ikhsanov-1998,
Ikhsanov-Beskrovnaya-2012}. Thus, the white dwarfs in the ejector state do exist but how
could they forme?

 \section{Accretion-driven spin-up}

As the surface temperature of the white dwarf in AE~Aquarii is limited
to\cite{Eracleous-etal-1994} $T_{\rm av} \leq 16\,000$\,K its
age\cite{Schoenberner-etal-2000} $t_{\rm cool} > 10^8$\,yr turns out to be significantly
larger than the spin-down timescale\cite{de-Jager-etal-1994} $t_{\rm sd} \leq 2 \times
10^7$\,yr. This indicates that the ejector white dwarf is a product of the binary
evolution which included an epoch of its rapid spin-up caused by intensive accretion.
The parameters of the spin-up epoch are the following\cite{Ikhsanov-Beskrovnaya-2012}: \\
i)~accretion from a Keplerian disk at a rate $\dot{M}_{\rm pe} \geq 10^{-7}\,{\rm M_{\sun}\,yr^{-1}}$
on a timescale of a few million years;\\
ii)~screening of the magnetic field of the white dwarf by the accreted material by a factor
of\cite{Bisnovatyi-Kogan-2006, Bisnovatyi-Kogan-Komberg-1974, Lovelace-etal-2005} 50--100;\\
iii)~re-emerging of the magnetic field due to diffusion of the stellar magnetic field
through the layer of the accreted material on a timescale $t_{\rm em} < t_{\rm sd}$,
where $t_{\rm sd} = P_{\rm s}/2 \dot{P}$ is the spin-down timescale which for the
parameters of AE~Aquarii is $\sim 2$\,Myr. For this condition to be satisfied the total
mass accreted by the white dwarf during the spin-up epoch should not exceed $\Delta
M_{\rm a} \leq 0.009\,M_{\sun}$ and the mass of the white dwarf itself is $M_{\rm wd}
\geq 1.1\,M_{\sun}$. This implies that the inclination angle of the binary system is
$\leq 54^{\degr}$.

An analysis of this accretion scenario suggests that the more massive the white dwarf is
the shorter is the spin period which it can reach at the end of the spin-up epoch.
Furthermore, the more massive a white dwarf is the smaller amount of the accreted
material is required to spin it up to the shortest possible spin period. The timescale
of the magnetic field re-emerging of the white dwarf after the spin-up epoch
is\cite{Cumming-2002} $\propto \Delta M_{\rm a}^{7/5}$. Therefore, the more massive the
white dwarf is the faster its magnetic field is re-generated. This increases a
probability for the white dwarf to appear as a star in the ejector state.

 \section*{Acknowledgments}

This research has been partly supported by the Presidium of RAS under the Program
Nr.\,21, Russian Ministry of Science and Education under the grants Nrs.\,8394 and 8417
and the Program ``National Scientific Schools'' under the grant NSH-1625.2012.2.

\end{document}